\documentclass[pre,aps,twocolumn,superscriptaddress]{revtex4-1}

\usepackage[utf8]{inputenc}
\usepackage{graphicx}
\usepackage{amssymb,amsfonts,amsmath}
\usepackage{color}
\usepackage[hidelinks]{hyperref}
\usepackage{bbm}
\usepackage{bbold}
\usepackage{pdfpages}
\makeatletter
\AtBeginDocument{\let\LS@rot\@undefined}
\makeatother
\usepackage{pgffor}

\usepackage{tikz}
\usetikzlibrary{shapes}
\usetikzlibrary{patterns}
\usetikzlibrary{angles, quotes}

\newcommand{\rv}{{\mathbf r}}

\newcommand{\Tr}{{\rm Tr}\,}
\newcommand{\e}{{\rm e}}

\newcommand{\pv}{{\bf p}}

\newcommand{\msphantom}[1]{$\ldots$}

\newcommand{\eqr}[1]{Eq.~\eqref{#1}}

\newcommand{\mydelete}[1]{{}}

\newcommand{\rmexc}{{\rm exc}}
\newcommand{\rmext}{{\rm ext}}

\newcommand{\rmid}{{\rm id}}
\newcommand{\x}{\rv}
\newcommand{\cov}{{\rm cov}}
\newcommand{\ofrn}{}
\newcommand{\chiA}{\chi_A}

\begin{document}

\title{Hyper-Density Functional Theory of Soft Matter}

\author{Florian Samm\"uller}
\affiliation{Theoretische Physik II, Physikalisches Institut, 
  Universit{\"a}t Bayreuth, D-95447 Bayreuth, Germany}

\author{Silas Robitschko}
\affiliation{Theoretische Physik II, Physikalisches Institut, 
  Universit{\"a}t Bayreuth, D-95447 Bayreuth, Germany}

\author{Sophie Hermann}
\affiliation{Theoretische Physik II, Physikalisches Institut, 
  Universit{\"a}t Bayreuth, D-95447 Bayreuth, Germany}

\author{Matthias Schmidt}
\affiliation{Theoretische Physik II, Physikalisches Institut, 
  Universit{\"a}t Bayreuth, D-95447 Bayreuth, Germany}
\email{Matthias.Schmidt@uni-bayreuth.de}

\date{12 March 2023, revised version: 25 July 2024}

\begin{abstract}
We present a scheme for investigating arbitrary thermal observables in
spatially inhomogeneous equilibrium many-body systems. Extending
the grand canonical ensemble yields any given observable as an
explicit hyper-density functional. Associated local fluctuation
profiles follow from an exact hyper-Ornstein-Zernike equation.
While the local compressibility and simple observables permit
analytic treatment, complex order parameters are accessible via
simulation-based supervised machine learning of neural hyper-direct
correlation functionals.  We exemplify efficient and accurate
neural predictions for the cluster statistics of hard rods, square
well rods, and hard spheres. The theory allows to treat complex
observables, as is impossible in standard density functional theory.
\end{abstract}

\maketitle

Classical density functional theory is a powerful framework for
describing the collective behaviour of a wide variety of relevant
many-body systems \cite{evans1979, evans1992, evans2016, hansen2013,
  schmidt2022rmp}.  Topical applications to soft matter
\cite{evans2019physicsToday} range from studies of hydrophobicity
\cite{levesque2012jcp, evans2015jpcm, evans2019pnas, coe2022prl,
  jeanmairet2013jcp} to investigations of the molecular structure of
liquids \cite{jeanmairet2013jcp,jeanmairet2013jpcl,
    luukkonen2020} and of electrolytes~\cite{martinjimenez2017natCom,
  hernandez-munoz2019, cats2021decay}.  The central variable of
density functional theory is the position-resolved one-body density
profile.  In recent developments, the local compressibility
\cite{evans2015jpcm, evans2019pnas, wilding2024} and more general
fluctuation profiles \cite{eckert2020, eckert2023fluctuation, coe2023}
were shown to be further useful indicators for collective phenomena,
e.g.\ when systematically analyzing drying that occurs near substrates
and around solutes \cite{evans2015jpcm, evans2019pnas, coe2023,
  wilding2024}. A further broad spectrum of observables, including
recent multi-body correlation functions \cite{zhang2020fourPoint,
  singh2023pnas, pihlajamaa2023}, are relevant for the study of
complex systems.

The use of statistical mechanical sum rules \cite{evans1979,
  hansen2013, evans1990, henderson1992} is often decisive in the
description of soft matter, as sum rules not only provide unambiguous
consistency checks, but also as they encapsulate physical constraints,
which ultimately facilitates to trace physical mechanisms and identify
underlying causes for the emerging collective effects. Inline with
further topical uses of the Noether theorem in Statistical Mechanics
\cite{revzen1970, marvian2014quantum, sasa2016, sasa2019, budkov2022,
  brandyshev2023, bravetti2023}, the recent thermal Noether invariance
theory \cite{hermann2021noether, hermann2022topicalReview,
  robitschko2024any} allows to systematically generate and classify a
significant body of exact sum rules. The properties of general
observables can be addressed via the recent hyperforce
theory~\cite{robitschko2024any}, which is similar in spirit to
Hirschfelder's hypervirial generalization \cite{hirschfelder1960} of
the standard virial theorem \cite{hansen2013}.

Machine learning techniques see a rapidly increasing use in soft
matter research across topics that range from characterization
\cite{clegg2021ml} to engineering of self-assembly
\cite{dijkstra2021ml}, detection of colloidal structure
\cite{boattinia2019ml}, and the study of effective colloidal
interaction potentials \cite{campos2021ml, campos2022ml}.  In the
context of density functional theory, machine learning was used for
the construction of workable representations for the central
functional both in the classical \cite{teixera2014, lin2019ml,
  lin2020ml, cats2022ml, qiao2020, yatsyshin2022, malpica-morales2023,
  fang2022, delasheras2023perspective, sammueller2023neural,
  sammueller2023whyNeural, dijkman2024ml,
  sammueller2024pairmatching, kelley2024ml} and in the quantum
realms \cite{nagai2018, jschmidt2018, zhou2019, nagai2020, li2021prl,
  li2022natcompsci, gedeon2022, pederson2022, kelley2024ml}. The
recent neural functional theory \cite{delasheras2023perspective,
  sammueller2023neural, sammueller2023whyNeural} constitutes a hybrid
method that is based on many-body computer simulation data used to
train a neural network, which then acts as a central numerical object
that mirrors the functional structure prescribed by classical density
functional theory.

Here we return to fundamentals and present a generalization of
classical density functional theory that allows to investigate the
behaviour of virtually arbitrary observables and their locally
resolved fluctuation profiles. We specifically develop a general
variational formalism based on an extended thermal ensemble
\cite{zwanzig2001eg, anero2013}. While the theory is
formally exact, we demonstrate as one way forward that the
relevant functional dependencies are amenable to supervised machine
learning. We present model demonstrations for what we argue is a
standalone and practically relevant computational scheme for the
investigation of soft matter. Our approach rests on the
Mermin-Evans functional map \cite{mermin1965, evans1979}, which
ascertains that any equilibrium quantity can be expressed as a
density functional.

Density functional theory \cite{evans1979,hansen2013} puts the
one-body density distribution,
\begin{align}
  \rho(\rv)=\langle\hat\rho(\rv)\rangle,
  \label{EQrhoAsAverage}
\end{align}
at center stage. Here we have defined the one-body density
``operator'' in its standard form $\hat\rho(\rv)=\sum_i
\delta(\rv-\rv_i)$, with $\delta(\cdot)$ denoting the Dirac
distribution, $\rv$ is a generic position variable, $\rv_i$ is the
position of particle $i=1,\ldots,N$, and $\langle \cdot \rangle$
indicates the thermal average as specified below in detail.  In the
following we consider a general observable that is represented by a
phase space function $\hat A(\rv^N)$, which in general depends on the
position coordinates~$\rv^N$ of all $N$ particles, as well as possibly
on additional parameters.  Following its occurrence in the hyperforce
theory \cite{robitschko2024any}, in generalization of the one-body
fluctuation profiles of Refs.~\cite{evans2015jpcm,
evans2019pnas,eckert2020, eckert2023fluctuation, coe2023}, and
borrowing the terminology from the hypervirial theorem
\cite{hirschfelder1960}, we consider a corresponding
hyper-fluctuation profile $\chi_A(\rv)$ as the covariance of the
density operator and the given observable. Together with the mean $A$
we hence define:
\begin{align}
  A &= \langle \hat A \rangle,
  \label{EQAmean}\\
  \chi_A(\rv) &=\cov(\hat\rho(\rv), \hat A\ofrn),
  \label{EQchiAsCovariance}
\end{align}
with the standard covariance $\cov(\hat X, \hat Y)= \langle \hat X
\hat Y \rangle - \langle \hat X\rangle \langle \hat Y \rangle$ of
two phase space functions $\hat X$ and $\hat Y$.

Specifically, we consider classical many-body systems of $N$
particles of identical mass $m$. The Hamiltonian has the standard form
$H=\sum_i \pv_i^2/(2m) + u(\rv^N) + \sum_i V_\rmext(\rv_i)$, where the
sums run over all particles $i=1,\ldots N$, the momentum of particle
$i$ is denoted by $\pv_i$, the interparticle interaction potential is
$u(\rv^N)$, where $\rv^N \equiv \rv_1,\ldots,\rv_N$ is a shorthand for
all position coordinates, and $V_\rmext(\rv)$ is an external
potential, here written as a function of the (generic) position
coordinate $\rv$. We work in the grand ensemble at absolute
temperature $T$ and chemical potential $\mu$.  The classical ``trace''
operation is defined as $\Tr \cdot = \sum_{N=0}^\infty (h^{dN}N!)^{-1}
\int d\rv^N d\pv^N \cdot$, where~$h$ denotes the Planck constant, $d$
the spatial dimensionality, and $\int d\rv^N d\pv^N$ indicates the
phase space integral over all position coordinates and momenta.

To incorporate the observable $\hat A(\rv^N)$ into the
framework, we consider an extended ensemble~\cite{zwanzig2001eg,anero2013}
that is here defined by the extended equilibrium
many-body probability distribution $\e^{-\beta(H-\mu N) + \lambda \hat
  A}/\Xi$, where $\lambda$ is a coupling parameter that acts as a
conjugate variable to $\hat A(\rv^N)$; here $\beta=1/(k_BT)$ with
$k_B$ denoting the Boltzmann constant. The extended grand partition
sum is given as $\Xi = \Tr \e^{-\beta (H-\mu N)+\lambda\hat A}$ and
the corresponding grand potential is $\Omega = -k_BT \ln\Xi$.  Thermal
averages are obtained via $\langle \cdot \rangle = \Tr \cdot
\e^{-\beta(H-\mu N) + \lambda \hat A}/\Xi$. An alternative and equivalent view of the
extended ensemble can be based on rather considering an extended
Hamiltonian $H_A = H - \lambda \hat A/\beta$ and formulating its
associated standard grand ensemble.

Despite the generalization we remain only interested in the
properties of the original system with Hamiltonian~$H$, recovering its
standard grand ensemble for the case of vanishing coupling constant,
$\lambda\to 0$. Throughout we assume that $\hat A(\rv^N)$ is of a form
such that the statistical ensemble generated via $H_A$ is well-defined
in this limit and that indeed $H_A\to H$ recovers the original
Hamiltonian (which specifies our above restriction to
``virtually'' arbitrary observables $\hat A$).

The thermal average~$A$ and the hyper-fluctuation profile
$\chiA(\rv)$, as respectively defined via Eqs.~\eqref{EQAmean} and
\eqref{EQchiAsCovariance}, are generated via the following partial
derivatives with respect to the coupling parameter $\lambda$:
\begin{align}
  A &= -\frac{\partial \beta\Omega}{\partial \lambda},
  \label{EQmeanAparametricDerivative}\\
  \chiA(\rv) &= \frac{\partial\rho(\rv)}{\partial\lambda}.
  \label{EQchiAsParametricDerivative}
\end{align}
The statepoint $\mu, T$ and the form of the external potential
$V_\rmext(\rv)$ are thereby fixed upon differentiating.  That
Eqs.~\eqref{EQmeanAparametricDerivative} and
\eqref{EQchiAsParametricDerivative} hold can respectively be verified
by elementary calculations taking into account that $-\beta
\Omega=\ln\Xi$ and the definition \eqref{EQrhoAsAverage} of
$\rho(\rv)$.  Equation \eqref{EQchiAsParametricDerivative} is also
rapidly derived from the standard expression
\cite{evans1979,hansen2013} of the density profile as a functional
derivative, $\delta \Omega /\delta V_\rmext(\rv) = \rho(\rv)$,
together with \eqr{EQchiAsCovariance} and the
recent~\cite{eckert2023fluctuation} general identity $-\delta A
/\delta \beta V_\rmext(\rv) = \cov(\hat\rho(\rv),\hat A)
=\chi_A(\rv)$. The latter relation is also straightforward to show by
explicit calculation and it lends much physical meaning to
$\chi_A(\rv)$ as the response function of the average $A$ against
changes in the shape of the scaled external potential $-\beta
V_\rmext(\rv)$.

The Euler-Lagrange equation of classical density functional theory
\cite{evans1979, evans1992, evans2016, hansen2013}, applied to the
extended Hamiltonian $H_A$, has the standard form:
\begin{align}
  c_1(\rv,[\rho]) = \ln\rho(\rv) + \beta V_\rmext(\rv) - \beta\mu,
  \label{EQel}
\end{align}
where $c_1(\rv,[\rho])$ is the one-body direct correlation functional
corresponding to $H_A$, i.e.\ for a system of particles that interact
via the extended interparticle interaction potential $u(\rv^N)-
\lambda \hat A(\rv^N)/\beta$.  In \eqr{EQel} we have set the thermal
de Broglie wavelength to unity and we denote functional dependence by
square brackets throughout. As \eqr{EQel} holds for any value of
$\lambda$, provided that $\rho(\rv)$ is the corresponding equilibrium
density profile, we can differentiate the equation with respect to
$\lambda$ and retain a valid identity. The result is the following
hyper-Ornstein-Zernike relation:
\begin{align}
  c_A(\rv,[\rho]) &=
  \frac{\chiA(\rv)}{\rho(\rv)} 
  -\int d\rv' c_2(\rv,\rv',[\rho]) \chiA(\rv'),
  \label{EQhyperOZ}
\end{align}
where $c_2(\rv,\rv',[\rho])=\delta c_1(\rv,[\rho])/\delta\rho(\rv')$
is the two-body direct correlation functional
\cite{evans1979, evans1992, evans2016, hansen2013}. The left hand
side of \eqr{EQhyperOZ} constitutes the hyper-direct correlation
functional $c_A(\rv,[\rho]$), as is obtained from parametrically
differentiating the left hand side of \eqr{EQel} at fixed density
profile:
\begin{align}
  c_A(\rv,[\rho]) &= 
  \frac{\partial c_1(\rv,[\rho])}{\partial \lambda} \Big|_{\rho}.
  \label{EQcAdefinition}
\end{align}

\begin{figure*}[htb!]
   \includegraphics[width=\textwidth]{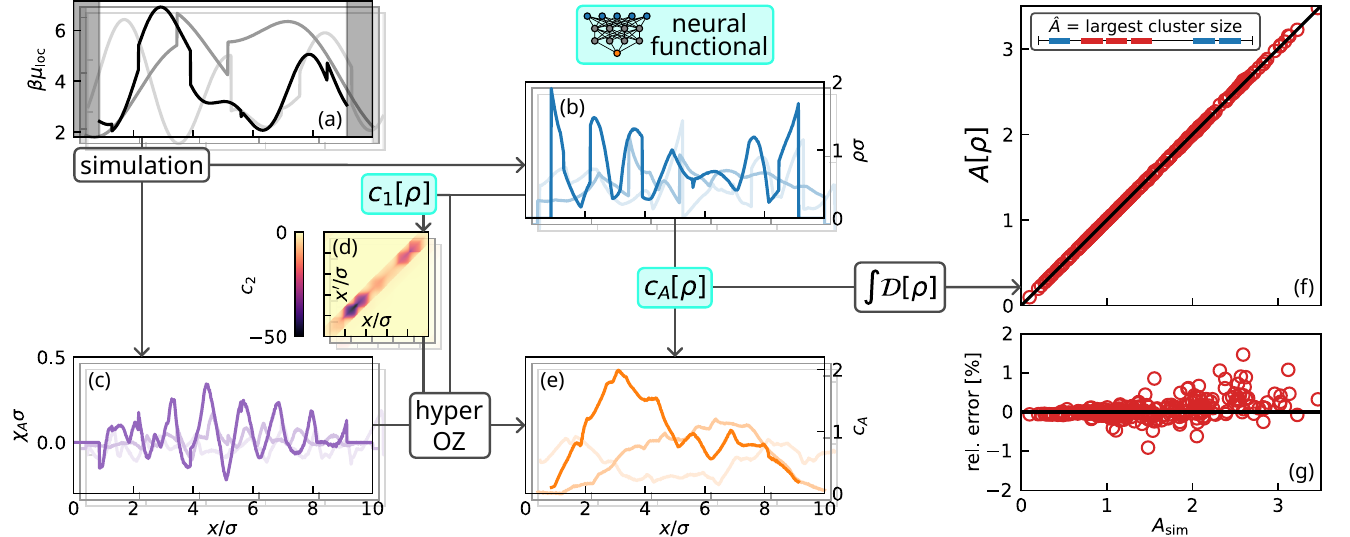}
   \caption{Overview of hyper-density functional theory.  The
     observable $\hat A$ is chosen as the number of particles
     belonging to the largest cluster in hard rods of size $\sigma$.
     (a) A local chemical potential $\beta\mu_{\rm
       loc}(x)=\beta\mu-\beta V_\rmext(x)$ creates spatially
     inhomogeneous systems. Shown are representative examples from 512
     grand canonical Monte Carlo simulations with both randomized
     values of $\beta\mu$ and forms of $\beta V_\rmext(x)$.  (b)
     Corresponding scaled density profiles $\rho(x)\sigma$ sampled via
     \eqr{EQrhoAsAverage}.  (c) Corresponding scaled hyper-fluctuation
     profiles $\chi_A(x)\sigma$ obtained via \eqr{EQchiAsCovariance}.
     (c) Two-body direct correlation function $c_2(x,x',[\rho])$, as
     obtained via autodiff from $c_1(x,[\rho])$
     \cite{sammueller2023neural, sammueller2023whyNeural}.  (e)
     Hyper-direct correlation functions $c_A(x)$ obtained by solving
     the hyper-Ornstein-Zernike equation \eqref{EQhyperOZ}.  Using the
     density profile as input and the simulation results for $c_A(x)$
     as target, supervised training yields a neural network that
     represents the hyper-direct correlation functional
     $c_A(x,[\rho])$.  (f) Predicted values $A[\rho]=\int D[\rho]
     c_A(x,[\rho])$ from functional integration according to
     \eqr{EQmeanAviaFunctionalIntegral}. For a test set of 256 systems
     not encountered during training the predictions of $A[\rho]$ are
     compared against reference simulation data~$A_{\rm sim}$.  (g)
     The relative numerical error of the predicted mean size $A$ of
     the largest cluster is smaller than $\sim 1\%$.
   \label{FIG1}
  }
\end{figure*}

It remains to express the thermal expectation value~$A$ via
relation~\eqref{EQmeanAparametricDerivative} as the
parametric derivative of the extended grand potential, taken while
keeping $\mu, T$ and the shape of $V_\rmext(\rv)$ fixed.  Crucially,
instead of considering explicit many-body expressions, we adopt the
classical density functional perspective \cite{evans1979, evans1992,
  evans2016, hansen2013, schmidt2022rmp} in order to find $A[\rho]$.
We hence work with the grand potential in its density functional form:
$\Omega[\rho]=F_\rmid[\rho]+F_\rmexc[\rho]+\int
d\rv\rho(\rv)[V_\rmext(\rv)-\mu]$.  Thereby the ideal gas free energy
functional is $F_\rmid[\rho]=k_BT \int d\rv \rho(\rv)[\ln\rho(\rv)-1]$
and $F_\rmexc[\rho]$ denotes the excess (over ideal gas) free energy
functional.

Differentiating $\Omega[\rho]$ with respect to $\lambda$ at fixed
$V_\rmext(\rv)$ gives one direct contribution and one term arising
from the induced changes in $\rho(\rv)$. The latter term, $\int d\rv
\delta \Omega[\rho]/\delta\rho(\rv)|_{V_\rmext}
\partial\rho(\rv)/\partial \lambda$, vanishes due to the stationarity
of the grand potential, $\delta
\Omega[\rho]/\delta\rho(\rv)|_{V_\rmext}=0$.  The direct contribution
is the derivative at fixed density, $ -\partial
\beta\Omega[\rho]/\partial \lambda |_{\rho}= -\partial \beta
F_\rmexc[\rho]/\partial\lambda|_{\rho}$, where we have exploited that
the ideal, external and chemical potential contributions to
$\Omega[\rho]$ are independent of $\lambda$. From recalling
\eqr{EQmeanAparametricDerivative} we obtain $A[\rho]=-\partial \beta
F_\rmexc[\rho]/\partial\lambda|_{\rho}$.

Functional calculus allows to re-write this formal expression
for $A[\rho]$ as a functional integral \cite{evans1979, evans1992,
  sammueller2023whyNeural}.  Using the standard relation
$c_1(\rv,[\rho])=-\delta\beta F_\rmexc[\rho]/\delta\rho(\rv)$ together
with the definition \eqref{EQcAdefinition} of the hyper-direct
correlation functional $c_A(\rv,[\rho])$ gives:
\begin{align}
  c_A(\rv,[\rho]) &= 
  \frac{\delta A[\rho]}{\delta\rho(\rv)},
  \label{EQcAasParametricDerivative}  \\
  A[\rho] &= \int \mathcal{D}[\rho] c_A(\rv,[\rho]).
  \label{EQmeanAviaFunctionalIntegral}
\end{align}
The functional derivative \eqr{EQcAasParametricDerivative} gives much
further significance to the hyper-direct correlation functional
$c_A(\rv,[\rho])$ as measuring changes of the thermal mean $A$ against
local perturbation of the density profile $\rho(\rv)$.
Equation~\eqref{EQmeanAviaFunctionalIntegral} is the inverse of
\eqr{EQcAasParametricDerivative} upon standard functional integration
\cite{evans1979, evans1992, sammueller2023whyNeural}. One can
efficiently parameterize the functional integral e.g.\ as $A[\rho] =
\int d\rv \rho(\rv) \int_0^1 da c_A(\rv,[a\rho])$, where the scaled
density profile $a\rho(\rv)$ is obtained by multiplication of
$\rho(\rv)$ with the parameter $0\leq a \leq 1$. Equation~\eqref{EQmeanAviaFunctionalIntegral}
allows to express the thermal
average of a given observable as an explicit density
functional~$A[\rho]$, provided that the density functional dependence
of $c_A(\rv,[\rho])$ is known.

As an initial test of this framework, we let the considered observable
simply be the particle number $\hat A(\rv^N)=N$, which we
recall is a fluctuating variable in the grand ensemble.
According to \eqr{EQchiAsCovariance} we have
$\chi_A(\rv)=\cov(\hat\rho(\rv), N)$ and from
\eqr{EQmeanAparametricDerivative} we obtain $N =-\partial \beta
\Omega/\partial \lambda$. Furthermore
\eqr{EQchiAsParametricDerivative} yields $\chi_A(\rv)=\partial
\rho(\rv)/\partial \lambda$. These are all key properties of the local
compressibility $\chi_\mu(\rv)=\beta\chi_A(\rv)$~\cite{evans2015jpcm,
  evans2019pnas, eckert2020, eckert2023fluctuation, coe2023,
  wilding2024} with the coupling parameter $\lambda$ playing the role
of the scaled chemical potential~$\beta \mu$. Hence for more general observables $\hat A$, we
conclude that $\chi_A(\rv)$ can be viewed as a corresponding
generalization of $\chi_\mu(\rv)$.

We next address significantly more complex forms of $\hat A$. We therefore return to
the hyper-Ornstein-Zernike equation \eqref{EQhyperOZ} and consider the
accessiblity of the terms on its right hand side on the basis of
direct simulations and the methods provided by the recent neural
functional theory \cite{sammueller2023neural, sammueller2023whyNeural}.
Both the standard density profile $\rho(\rv)$ and the
hyper-fluctuation profile $\chiA(\rv)$ can be sampled for given $\mu,
T$, and $V_\rmext(\rv)$, recall $\rho(\rv)$ as the average
\eqref{EQrhoAsAverage} and $\chiA(\rv)$ as the
covariance~\eqref{EQchiAsCovariance}.  For the given bare
Hamiltonian~$H$, the neural functional theory allows to construct a
neural-network-based representation of the direct correlation functional
$c_1(\rv,[\rho])$ \cite{sammueller2023neural,
  sammueller2023whyNeural}.  Automatic differentiation then
straightforwardly provides a numerically efficient and accurate neural
functional representation of
$c_2(\rv,\rv',[\rho])$ which is ready for use in \eqr{EQhyperOZ}.
Evaluating the right hand side of \eqr{EQhyperOZ} then only requires
the numerical integration over~$\rv'$.

\begin{figure}[t]
  \includegraphics[width=0.99\columnwidth]{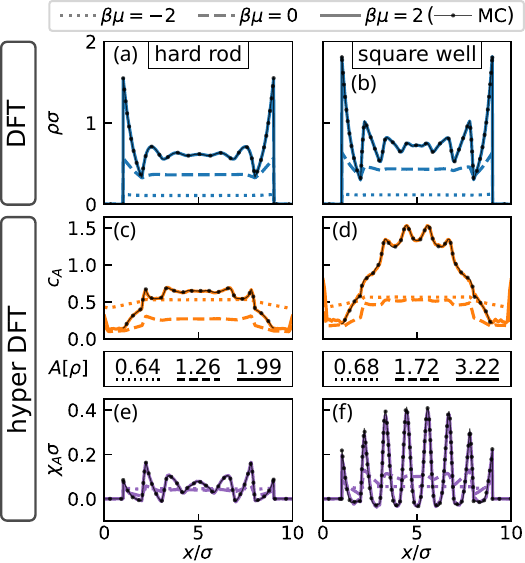}
  \caption{Application of the hyper-density functional theory to the
    statistics of the largest cluster of hard rods [(a), (c), (e)] and
    attractive square-well rods [(b), (d), (f)].  The systems are
    confined between two hard walls with separation distance
    $L=8\sigma$ and the profiles are shown as a function of the scaled
    distance $x/\sigma$ across the pore for $\beta\mu=-2$ (dotted), 0
    (dashed), and 2 (solid lines).  We depict theoretical results for
    the density profile $\rho(x)\sigma$ [(a), (b)], the
    hyper-direct correlation function $c_A(x)$ [(c), (d)], the
    mean value $A[\rho]$, and the hyper-fluctuation profile
    $\chi_A(x)\sigma$ [(e), (f)]. For the case of largest
    chemical potential, Monte Carlo simulation results are shown as
    reference (solid lines with symbols). These coincide with
    the respective theoretical predictions on the scale of the
    plot. The simulated reference values for $A$ are identical within
    the quoted accuracy to the neural predictions $A[\rho]$.
    \label{FIG2}}
\end{figure}

Hence, for given $\mu, T$, and $V_\rmext(\rv)$, the hyper-direct
correlation function $c_A(\rv)$ that is specific for the considered
inhomogeneous system can be computed explicitly via \eqr{EQhyperOZ}. This facilitates the generation
of a training data set from many-body simulation results.  We do not
invoke the functional dependence of $c_A(\rv,[\rho])$ for this task
yet and require only standard grand canonical Monte Carlo simulation
techniques \cite{frenkel2023book, wilding2001, brukhno2021dlmonte}
with no need to implement the extended ensemble explicitly.

Following the neural methodology for standard density
functionals~\cite{sammueller2023neural,sammueller2023whyNeural}, this
puts us in the position to machine learn
the hyper-functional map
\begin{align}
  \rho(\rv') \to c_A(\rv),
  \label{EQrhoTocA} 
\end{align}
where the density at positions $\rv'$ removed from $\rv$ will
contribute with a range of non-locality that is specific to the form
of the observable $\hat A(\rv^N)$. We proceed in analogy to
Refs.~\cite{sammueller2023neural, sammueller2023whyNeural} in
constructing a neural network representation of $c_A(\rv,[\rho])$ via
supervised machine learning on the basis of randomized training data
sets, where at fixed temperature $T$, the value of $\mu$ and the shape
of $V_\rmext(\rv)$ are varied. An illustration of the principal
workflow is shown in Fig.~1 as applied to the following physical
setup.

We first choose one-dimensional systems with either pure hard
core interactions or additional square well attraction. We
solely work on the basis of neural functionals to demonstrate the
independence from the availability of analytic free energy
functionals.  We have checked that using Percus' exact free energy
functional for hard rods \cite{percus1976, robledo1981} instead of
its neural representation~\cite{sammueller2023whyNeural}
generates no relevant numerical changes. The density profile under
the influence of an external potential follows from numerical solution
of the Euler-Lagrange equation~\eqref{EQel} for $\lambda=0$, using the
respective neural one-body direct correlation functional. As an order
parameter with genuine many-body character, we investigate cluster
properties and therefore define two particles $i$ and $j$ as bonded if
their positions $x_i$ and $x_j$ are within a bonding
cutoff, $|x_i-x_j|<x_c$, where we choose $x_c=1.2\sigma$, with the
hard-core particle diameter $\sigma$. For each microstate of positions
$x^N=x_1,\ldots,x_N$, we construct an instantaneous histogram that
gives the number of clusters with (integer) size $m$, where a cluster
consists of all particles that are bonded directly or mediated by
other bonded particles. This is a standard criterion that is
independent of dimensionality and used in studies of gelation
\cite{jadrich2023preface,sammueller2023gel}.

Specifically we choose $\hat A$ as the size of the largest cluster
in a given microstate. We find studying and
comparing the behaviour of hard-core and square-well attractive
rods to be a crucial test, as there is no way to assess the
respective clustering properties via conventional density
functional methods. Exemplary profiles are shown in Fig.~\ref{FIG1}
together with the numerical predictions from the neural functional
$A[\rho]$, as evaluated via the functional integral
\eqref{EQmeanAviaFunctionalIntegral}. The neural predictions are
highly accurate with a relative error consistently below~$\sim 1\%$ as
compared to the simulation reference.

On the basis of the availability of the neural hyper-direct
correlation functional we can formulate a template for standalone
application of the hyper-density functional theory to predict inhomogeneous states of a given fluid model, with no reliance
on further simulation data. We require trained neural network
representations for $c_A(\rv,[\rho])$ and $c_1(\rv,[\rho])$.
Automatic functional differentiation of the latter yields a neural
representation of $c_2(\rv,\rv',[\rho])$.

First, in a conventional density functional setting the solution of
the Euler-Lagrange equation \eqref{EQel} at given $\mu, T$ and $V_\rmext(\rv)$ yields the
shape of the equilibrium density profile~$\rho(\rv)$. This form is
then used to evaluate the hyper-direct correlation functional
$c_A(\rv,[\rho])$ as well as $c_2(\rv,\rv',[\rho])$. The resulting
functions turn the hyper-Ornstein-Zernike relation~\eqref{EQhyperOZ}
into a concrete integral equation for determining the
hyper-fluctuation profile $\chi_A(\rv)$. Predictions for the mean
value $A$ in the considered system are obtained from calculating
$A[\rho]$ at the known equilibrium density profile via numerical
functional integration~\eqref{EQmeanAviaFunctionalIntegral}.

Figure \ref{FIG2} shows results from this strategy applied to the
cluster statistics of both the hard rod and square-well system
(potential range $0.2\sigma$ and depth
$\beta\epsilon=1$)~\cite{sammueller2024hyperDFTgithub}.  We consider
confinement between two hard walls, but with no further disturbing
influence as was present during training. This clean situation tests
the genuine extrapolation capability of the neural functionals. The
results shown in Fig.~\ref{FIG2} achieve excellent agreement with
reference simulation data. This successful application demonstrates
that we have developed a systematic functional approach that allows to
address the equilibrium behaviour of general observables. The
statistical mechanical many-body problem is thereby cast into
functional form, which we have shown to be accessible via
simulation-based training of neural networks that can be applied
efficiently in predictive tasks.

We demonstrate that the demands for training the neural functionals
$c_1(\rv,[\rho])$ and $c_A(\rv,[\rho])$ are not prohibitive for
studying complex order parameters in realistic fluid models by
successful application of the cluster analysis to the hard sphere
system in three dimensions, as shown in our
SI~\cite{sammueller2024hyperDFTsupportingInformation}.  As all
quantities in the framework have physical interpretations, with
$\chi_A(\rv)$ generalizing the prominently used local compressibility
$\chi_\mu(\rv)$ \cite{evans2015jpcm, evans2019pnas, wilding2024,
  eckert2020,eckert2023fluctuation, coe2023} and thermal
susceptibility $\chi_T(\rv)$ \cite{eckert2020,eckert2023fluctuation,
  coe2023}, it seems feasible that existing density
functional methods \cite{evans1979, evans1992, evans2016, hansen2013}
could be applied in analytical hyper-functional construction. The ease
with which neural hyper-functionals can be trained can motivate such
work.  Analytical treatments of simple one- and two-body forms of
$\hat A(\rv^N)$, the relationships to a static version of the
countoscope \cite{mackay2023countoscope} and to force-density
functional theory \cite{tschopp2022forceDFT}, along with a
description of the role of $\chi_A(\rv)$ as the local compressibility and in the hyperforce
theory~\cite{robitschko2024any} are given in our
SI~\cite{sammueller2024hyperDFTsupportingInformation}.  While we
have focused on dependence on position, we see no formal problems in
incorporating orientational degrees of freedom. The increase in
complexity could be alleviated by the use of molecular density
functional methods \cite{jeanmairet2013jpcl, jeanmairet2013jcp,
  luukkonen2020}.  Investigating deeper relationships with
functional thermo-dynamics \cite{anero2013}, thermal Noether
invariance \cite{hermann2021noether, hermann2022topicalReview,
  robitschko2024any}, and power functional
theory~\cite{schmidt2022rmp} is worthwhile.



\section*{Supplementary material}

# Supplemental Material for:
# Hyper-Density Functional Theory of Soft Matter

Florian Sammüller,[1] Silas Robitschko,[1] Sophie Hermann,[1] and Matthias Schmidt[1]

[1]*Theoretische Physik II, Physikalisches Institut, Universität Bayreuth, D-95447 Bayreuth, Germany*

(Dated: 25 July 2024)

## I. NEURAL CLUSTER HYPER-DENSITY FUNCTIONAL FOR HARD SPHERES

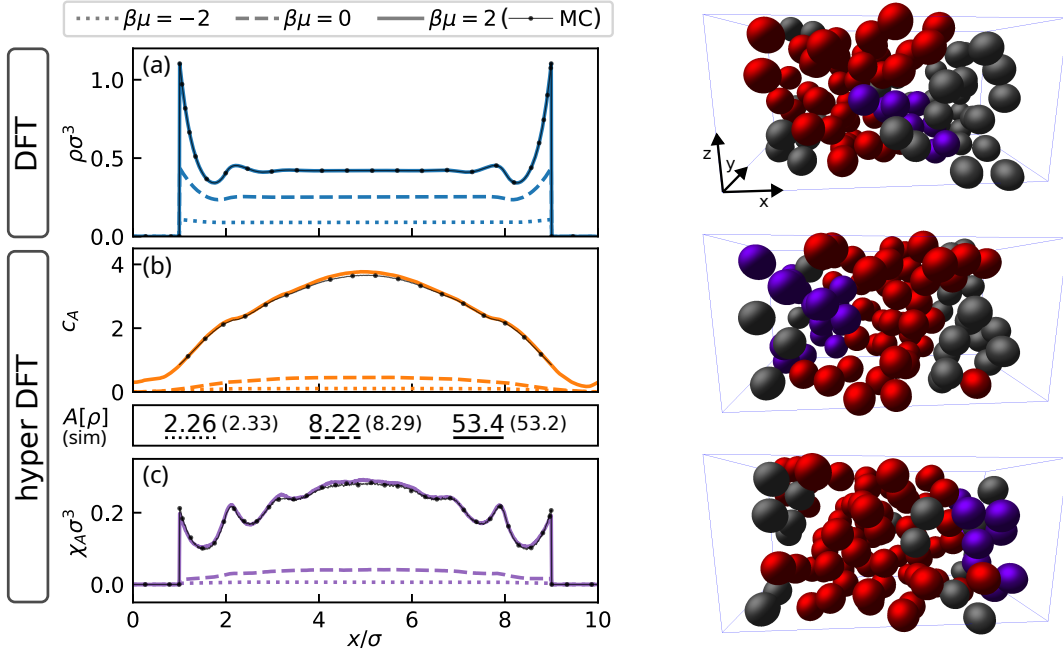

FIG. S1. Hyper-density functional results for the cluster statistics of three-dimensional hard spheres. The order parameter $\hat{A}$ is chosen as the number of particles in the largest cluster. The system is under planar confinement between two hard walls with separation distance $8\sigma$. Theoretical results are shown for three different values of the scaled chemical potential: $\beta\mu = -2$ (dotted), 0 (dashed), and 2 (solid lines); reference simulation data is given for the latter case $\beta\mu = 2$ (symbols). (a) Shown are neural density functional results for the scaled density profile $\rho(x)\sigma^3$ as a function of the scaled distance $x/\sigma$ across the slit. The results are obtained from numerical solution of the Euler-Lagrange equation of classical density functional theory, using the neural hard sphere one-body correlation functional $c_1(x,[\rho])$ of Ref. [2] as input. (b) Corresponding hyper-direct correlation functions $c_A(x)$ are obtained from evaluating the trained neural hyper-direct correlation functional $c_A(x,[\rho])$ at the three respective density profiles. Functional integration according to $A[\rho] = \int \mathcal{D}[\rho] c_A(x,[\rho])$ [see Eq. (10) in the main text] gives theoretical predictions for the mean value $A$ of the size of the largest cluster, as compared to the simulation reference $A = \langle \hat{A} \rangle$ (values in parenthesis). (c) Theoretical results for hyper-fluctuation profiles $\chi_A(x)$ are obtained from solving the hyper-direct Ornstein-Zernike relation [Eq. (7) in the main text] for the three considered situations using as input $\rho(x)$ and the neural functionals $c_A(x,[\rho])$ and $c_2(x,x',[\rho]) = \delta c_1(x,[\rho])/\delta\rho(x')$. The simulation reference for $\chi_A(x)$ is obtained from sampling the covariance $\chi_A(x) = \mathrm{cov}(\hat{\rho}(x), \hat{A})$ [see Eq. (3) in the main text]. The three simulation snapshots (right column) depict typical confined hard sphere configurations for $\beta\mu = 2$. The highlighted particles belong to the largest cluster (bright red) or to the second-largest cluster (dark violet). The number $\hat{A}$ of particles in the largest cluster fluctuates considerably over microstates.

We demonstrate the applicability of the neural hyper-density functional theory to more realistic fluid models than the one-dimensional systems considered in the main text. We address the mean cluster statistics of three-dimensional hard spheres based on neural functional methods. The observable $\hat{A}$ is the number of particles in the largest cluster, as described in the main text. We restrict ourselves to planar geometry and consider systems that are periodic in the two lateral directions with overall simulation box size $10\sigma \times 5\sigma \times 5\sigma$ [1]. This geometric reduction extends to all considered quantities and equations, including the hyper-Ornstein-Zernike equation [Eq. (7) in the main text]. The supervised training of $c_A(x,[\rho])$ is based on 428 simulation runs under randomized conditions both of the scaled

chemical potential $\beta\mu \in [-5, 10]$ and the form of the external potential $V_{\text{ext}}(x)$ [1, 2]. As a representative test case not encountered during training, we consider confinement between two hard walls with separation distance $8\sigma$. As shown in Fig. S1 the hyper-density functional theory yields results with near-simulation precision.

## II. ONE-BODY AND TWO-BODY OBSERVABLES

We address observables $\hat{A}(\mathbf{r}^N)$ of simple one- and two-body form that reduce the hyper-density functional approach to standard density functional settings. We first consider cases where the observable $\hat{A}(\mathbf{r}^N)$ is of one-body form

$$\hat{A}(\mathbf{r}^N) = \sum_i a_1(\mathbf{r}_i), \tag{S1}$$

where $a_1(\mathbf{r})$ is a given function of position $\mathbf{r}$. Using the one-body density operator $\hat{\rho}(\mathbf{r}) = \sum_i \delta(\mathbf{r} - \mathbf{r}_i)$, we can rewrite Eq. (S1) as $\hat{A}(\mathbf{r}^N) = \int d\mathbf{r} \hat{\rho}(\mathbf{r}) a_1(\mathbf{r})$. Building the equilibrium average thereof yields the thermal mean $A$ as a density functional

$$A[\rho] = \int d\mathbf{r} \rho(\mathbf{r}) a_1(\mathbf{r}). \tag{S2}$$

The corresponding hyper-direct correlation functional $c_A(\mathbf{r}, [\rho])$ follows from $c_A(\mathbf{r}, [\rho]) = \delta A[\rho]/\delta\rho(\mathbf{r})$ [Eq. (9) in the main text] as

$$c_A(\mathbf{r}, [\rho]) = a_1(\mathbf{r}). \tag{S3}$$

The associated hyper-fluctuation profile $\chi_A(\mathbf{r})$ follows from the covariance form $\chi_A(\mathbf{r}) = \text{cov}(\hat{\rho}(\mathbf{r}), \hat{A})$ [Eq. (3) in the main text] as

$$\chi_A(\mathbf{r}) = \int d\mathbf{r}' a_1(\mathbf{r}') \text{cov}(\hat{\rho}(\mathbf{r}), \hat{\rho}(\mathbf{r}')) = \int d\mathbf{r}' a_1(\mathbf{r}') H_2(\mathbf{r}, \mathbf{r}'), \tag{S4}$$

where $H_2(\mathbf{r}, \mathbf{r}') = \text{cov}(\hat{\rho}(\mathbf{r}), \hat{\rho}(\mathbf{r}'))$ is the standard density-density correlation function [3, 4]. Inserting Eqs. (S3) and (S4) into the hyper-Ornstein-Zernike equation $c_A(\mathbf{r}, [\rho]) = \chi_A(\mathbf{r})/\rho(\mathbf{r}) - \int d\mathbf{r}' c_2(\mathbf{r}, \mathbf{r}', [\rho]) \chi_A(\mathbf{r}')$ [Eq. (7) in the main text] yields the specific one-body form of the hyper-Ornstein-Zernike equation

$$\int d\mathbf{r}'' a_1(\mathbf{r}'') \Big[ \rho(\mathbf{r}) \int d\mathbf{r}' c_2(\mathbf{r}, \mathbf{r}') H_2(\mathbf{r}', \mathbf{r}'') + \rho(\mathbf{r}) \delta(\mathbf{r} - \mathbf{r}'') - H_2(\mathbf{r}, \mathbf{r}'') \Big] = 0. \tag{S5}$$

The identity (S5) is straightforward to prove directly as the term in brackets already vanishes due to the standard inhomogeneous two-body Ornstein-Zernike equation [3, 5] $H_2(\mathbf{r}, \mathbf{r}') = \rho(\mathbf{r})\delta(\mathbf{r} - \mathbf{r}') + \rho(\mathbf{r}) \int d\mathbf{r}'' c_2(\mathbf{r}, \mathbf{r}'') H_2(\mathbf{r}'', \mathbf{r}')$.

In alternative reasoning, Eq. (S5) holds for any permissible form of $a_1(\mathbf{r})$. Hence a valid identity is retained upon functionally differentiating Eq. (S5) with respect to $a_1(\mathbf{r})$ or, analogously, treating $a_1(\mathbf{r})$ as a mere test function. Both routes give the standard inhomogeneous two-body Ornstein-Zernike relation starting from the hyper-Ornstein-Zernike relation (S5), which serves as a consistency check.

For the specific choice $\hat{A} = N$, i.e. $a_1(\mathbf{r}) = 1$ in Eq. (S1), the hyper-functional framework relates back to the local compressibility $\chi_\mu(\mathbf{r}) = \beta\chi_A(\mathbf{r})$, which is a prominently featured fluctuation profile [6, 7, 9–12]. Turning to the density functional dependence, the hyper-direct correlation functional becomes constant unity, $c_A(\mathbf{r}, [\rho]) = 1$, in accordance with Eq. (S3). As a consequence the hyper-Ornstein-Zernike equation [Eq. (7) in the main text] reduces upon multiplication by $\rho(\mathbf{r})$ to $\rho(\mathbf{r}) \int d\mathbf{r}' c_2(\mathbf{r}, \mathbf{r}', [\rho]) \chi_\mu(\mathbf{r}') + \beta\rho(\mathbf{r}) = \chi_\mu(\mathbf{r})$, which is the fluctuation Ornstein-Zernike relation [10, 11] for the local compressibility $\chi_\mu(\mathbf{r})$. The functional integral [Eq. (10) in the main text] reduces according to Eq. (S2) to the explicit result $A[\rho] = \int d\mathbf{r}\rho(\mathbf{r})$. Hence the mean number of particles, expressed as the density functional $\bar{N}[\rho] = \int d\mathbf{r}\rho(\mathbf{r})$, is recovered, which certainly is correct due to $\int d\mathbf{r}\rho(\mathbf{r}) = \int d\mathbf{r}\langle\hat{\rho}(\mathbf{r})\rangle = \langle\int d\mathbf{r}\hat{\rho}(\mathbf{r})\rangle = \langle N \rangle = A$.

The reduction of the hyper-density functional framework to spatial integration of the standard Ornstein-Zernike relation extends to two-body forms of $\hat{A}(\mathbf{r}^N)$ albeit with an increase in complexity. We sketch essentials of the two-body case, where the observable under consideration has the form

$$\hat{A}(\mathbf{r}^N) = \sum_{ij} a_2(\mathbf{r}_i, \mathbf{r}_j), \tag{S6}$$

with the sums over $i$ and $j$ running over all particles and the function $a_2(\mathbf{r}, \mathbf{r}')$ is a given two-body field that depends on positions $\mathbf{r}$ and $\mathbf{r}'$. Using two density operators $\hat{\rho}(\mathbf{r})$ and $\hat{\rho}(\mathbf{r}')$ we can re-write Eq. (S6) as $\hat{A}(\mathbf{r}^N) =$

$\int d\mathbf{r} d\mathbf{r}' \hat{\rho}(\mathbf{r}) \hat{\rho}(\mathbf{r}') a_2(\mathbf{r}, \mathbf{r}')$. In order to formulate the mean $A$, we build the thermal average on both sides. Observing that $\langle \hat{\rho}(\mathbf{r}) \hat{\rho}(\mathbf{r}') \rangle = H_2(\mathbf{r}, \mathbf{r}') + \rho(\mathbf{r})\rho(\mathbf{r}')$ the result is

$$A[\rho] = \int d\mathbf{r} d\mathbf{r}' \rho(\mathbf{r}) \rho(\mathbf{r}') a_2(\mathbf{r}, \mathbf{r}') + \int d\mathbf{r} d\mathbf{r}' H_2(\mathbf{r}, \mathbf{r}') a_2(\mathbf{r}, \mathbf{r}'), \tag{S7}$$

where the first term on the right hand side is already an explicit density functional. Expressing the second term also as a density functional requires to have access to $H_2(\mathbf{r}, \mathbf{r}', [\rho])$ as a density functional, which is nontrivial. Conventionally one would base this functional dependence on solving the inhomogeneous two-body Ornstein-Zernike equation for the specific situation at hand and having access to $c_2(\mathbf{r}, \mathbf{r}', [\rho]) = -\delta^2 \beta F_{\text{exc}}[\rho]/[\delta \rho(\mathbf{r}) \delta \rho(\mathbf{r}')]$ as a density functional (see, e.g., Ref. [13] and references therein), thus providing a link to the standard theory of inhomogeneous liquids.

## III. RELATIONSHIP TO HYPERFORCE CORRELATIONS

Besides its present role in the hyper-density functional approach, the hyper-fluctuation profile $\chi_A(\mathbf{r})$ features prominently in the recently formulated exact hyperforce sum rule [14]

$$-\nabla \chi_A(\mathbf{r}) - \chi_A(\mathbf{r}) \nabla \beta V_{\text{ext}}(\mathbf{r}) + \text{cov}(\hat{A}(\mathbf{r}^N), \beta \hat{\mathbf{F}}_{\text{int}}(\mathbf{r})) + \left\langle \sum_i \delta(\mathbf{r} - \mathbf{r}_i) \nabla_i \hat{A}(\mathbf{r}^N) \right\rangle = 0, \tag{S8}$$

where $\hat{\mathbf{F}}_{\text{int}}(\mathbf{r}) = -\sum_i \delta(\mathbf{r} - \mathbf{r}_i) \nabla_i u(\mathbf{r}^N)$ is the one-body interparticle force density operator and $\hat{A}(\mathbf{r}^N)$ can have general many-body dependence on the coordinates $\mathbf{r}^N$. The form (S8) is straightforwardly obtained from starting with $\langle \hat{A} \beta \hat{\mathbf{F}}(\mathbf{r}) \rangle + \langle \sum_i \delta(\mathbf{r} - \mathbf{r}_i) \nabla_i \hat{A}(\mathbf{r}^N) \rangle = 0$ [14] and substracting the vanishing contribution $\langle \beta \hat{\mathbf{F}}(\mathbf{r}) \rangle \langle \hat{A} \rangle = 0$, which is zero due to $\langle \hat{\mathbf{F}}(\mathbf{r}) \rangle = 0$, hence allowing to convert the correlation expression (mean of the product) into covariance form. Then expressing $\hat{\mathbf{F}}(\mathbf{r})$ via its three constituent kinetic, external, and interparticle contributions, $\hat{\mathbf{F}}(\mathbf{r}) = \nabla \cdot \hat{\boldsymbol{\tau}}(\mathbf{r}) - \hat{\rho}(\mathbf{r}) \nabla V_{\text{ext}}(\mathbf{r}) + \hat{\mathbf{F}}_{\text{int}}(\mathbf{r})$, where the kinetic stress phase space function is $\hat{\boldsymbol{\tau}}(\mathbf{r}) = -\sum_i \delta(\mathbf{r} - \mathbf{r}_i) \mathbf{p}_i \mathbf{p}_i / m$, yields Eq. (S8). The sum rule (S8) is derived on the basis of the standard ensemble [14] and hence all averages correspond to the original Hamiltonian $H$. Nevertheless, the hyper-fluctuation profile $\chi_A(\mathbf{r})$ emerges naturally in the derivation based on the thermal invariance against phase space shifting [14].

For one-body observables (S1) we can find a simple link to Eq. (S8). We address the last term on the right hand side thereof as $\langle \sum_i \delta(\mathbf{r} - \mathbf{r}_i) \nabla_i \hat{A}(\mathbf{r}^N) \rangle = \langle \sum_i \delta(\mathbf{r} - \mathbf{r}_i) \nabla_i a_1(\mathbf{r}_i) \rangle = \rho(\mathbf{r}) \nabla a_1(\mathbf{r}) = \rho(\mathbf{r}) \nabla c_A(\mathbf{r})$, where in the first step we have started from $\hat{A}(\mathbf{r}^N) = \sum_i a_1(\mathbf{r}_i)$, which gives $\nabla_i \hat{A}(\mathbf{r}^N) = \nabla_i a(\mathbf{r}_i)$, and then have identified the specific form $c_A(\mathbf{r}) = a_1(\mathbf{r})$ according to Eq. (S3). Hence for one-body forms (S1) we have shown that $\langle \sum_i \delta(\mathbf{r} - \mathbf{r}_i) \nabla_i \hat{A}(\mathbf{r}^N) \rangle = \rho(\mathbf{r}) \nabla c_A(\mathbf{r}, [\rho])$, which is a simple limiting case for the behaviour of $c_A(\mathbf{r}, [\rho])$.

A specific example for a one-body observable $\hat{A}$ could be a simple model of a static countoscope [15], where $a_1(\mathbf{r}) = \Theta(l - |\mathbf{r}|)$ is an indicator function for the countoscope of diameter $2l$; here $\Theta(\cdot)$ indicates the Heaviside unit step function. Alternatively, one could use a version with planar symmetry, $a_1(\mathbf{r}) = \Theta(l - |x|)$, where $x$ is a Cartesian coordinate of $\mathbf{r}$.

Two-body observables $\hat{A}$ of the form (S6) arise naturally in systems governed by pairwise interparticle interactions, where the interparticle potential energy can be written as $u(\mathbf{r}^N) = \sum_{ij, i \neq j} \phi(|\mathbf{r}_i - \mathbf{r}_j|)/2$; see Ref. [13] for the formulation of classical density functional theory from the force point of view, which is based on expressing the local force density $\langle \hat{\mathbf{F}}_{\text{int}}(\mathbf{r}) \rangle$ as a two-body observable when only pair interparticle interactions are present in the system.

In general, the hyper-density functional theory allows the treatment of both many-body observables $\hat{A}$ and many-body interparticle interaction potentials $u(\mathbf{r}^N)$. The simple cases laid out here can act as a reference for making future analytical progress in these directions.

---



\end{document}